\documentclass[fleqn,10pt]{wlscirep}
\usepackage[utf8]{inputenc}
\usepackage[T1]{fontenc}
\usepackage{bm}

\title{Oscillatory active microrheology of active suspensions}

\author[1,*]{Milo\v s Kne\v zevi\'c}
\author[1]{Luisa E. Avil\'es Podgurski}
\author[1]{Holger Stark}
\affil[1]{Institut f\"ur Theoretische Physik, Technische Universit\"at Berlin, Hardenbergstra\ss e 36, D-10623 Berlin, Germany}

\affil[*]{knezevic@campus.tu-berlin.de}

\begin{abstract}
Using the method of Brownian dynamics, we investigate the dynamic properties of a 2d suspension of active disks at high P\'eclet numbers 
using active microrheology. In our simulations the tracer particle is driven either by a constant or an oscillatory external force. In the first case, we find that the mobility of the tracer initially appreciably decreases with the external force and then becomes approximately constant for larger forces. For an oscillatory driving force we find that the dynamic mobility shows a quite complex behavior -- it displays a highly nonlinear behavior on both the amplitude and frequency of the driving force. This result is important because it reveals that a phenomenological description 
of tracer motion in active media in terms of a simple linear stochastic equation even with a memory-mobility kernel is not appropriate.
\end{abstract}
\begin{document}

\flushbottom
\maketitle
\thispagestyle{empty}

\section*{Introduction}

\begin{figure}[ht]
	\centering
	\includegraphics[width=8cm]{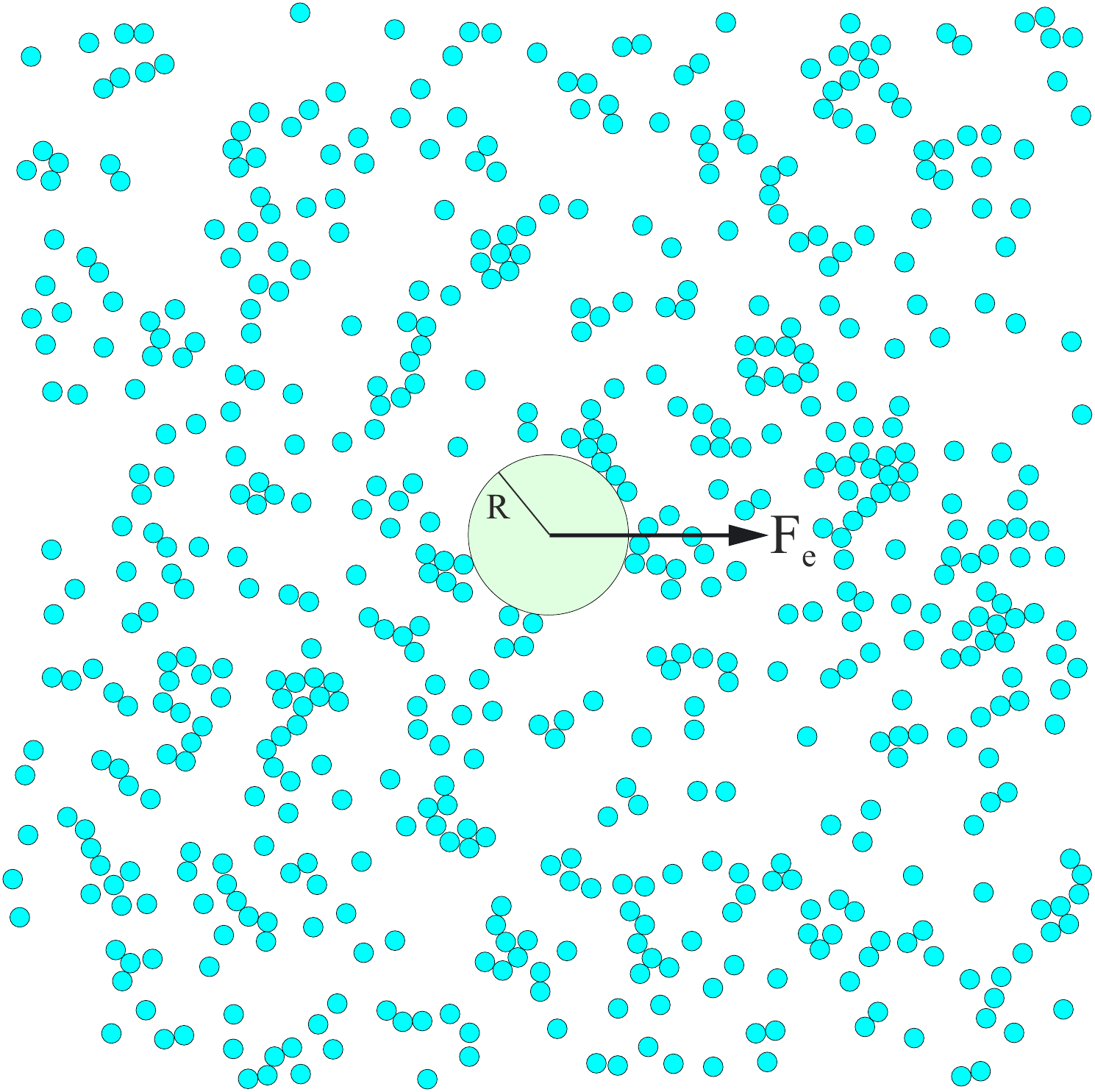}
	\caption{A snapshot of a passive tracer of radius $R$ driven by an external force $\mathbf{F}_\mathrm{e}(t) = F_\mathrm{e}(t)\mathbf{e}_x$ through a bath consisting of active disks.}
	\label{fig:1}
\end{figure} 

In the past three decades microrheology\cite{mason:95,mason:96,gittes:97,schnurr:97} has emerged as a useful technique for characterizing rheological properties of complex fluids, i.e. fluids that incorporate mesoscopic building units such as colloids, polymers or more complicated self-assembled structures. Unlike traditional rheology\cite{larson:99} which quantifies the viscoelastic properties of complex fluids by connecting stresses and (rates of) strain in the medium, in a typical microrheological\cite{mackintosh:99,waigh:05,cicuta:07,chen:10,wilson:11,waigh:16,zia:18} measurement one follows the trajectory of a probe particle (tracer) embedded in the medium under consideration. A microrheological study can be conducted in two ways: passive and active. Passive microrheology, in which one tracks the random motion of the tracer induced by fluctuations in the medium, has been employed to study a broad assortment of complex systems, spanning from colloidal suspensions\cite{carpen:05,khair:05,puertas:14} to biological matter\cite{weihs:06,wirtz:09}. In active microrheology, one either measures the mobility of the tracer particle driven by a static force\cite{squires:05,khair:06,zia:10}, or probes the frequency dependence of mobility by subjecting the tracer to an oscillatory external driving
force\cite{swan:14}. Additional information about the material's response can be acquired by utilizing rotational microrheology of anisotropic probes\cite{schmiedeberg:05}.

In this article we investigate the active microrheological response of a low Reynolds number suspension of active disks exhibiting self-propulsive motion\cite{romanczuk:12,elgeti:15,zoettl:16,bechinger:16}. They expend energy to propel themselves forward, and therefore constitute a nonequilibrium suspension, which displays a plethora of intriguing properties. For example, active particles can exert an active pressure on confining surfaces\cite{takatori:14,solon:15a,solon:15b,zakine:20,malgaretti:21} while getting stranded at them\cite{elgeti:09,schaar:15}, which can be exploited for constructing rotational\cite{angelani:09,dileonardo:10,sokolov:10} and translational\cite{angelani:10,kaiser:14,mallory:14,knezevic:20} ratchet motors powered by active particles as well as for cargo transport\cite{palacci:13,koumakis:13} and active self-organization\cite{simmchen:16,stenhammar:16,mallory:18}. Despite these dazzling features of active suspensions, their microrheological properties are still not sufficiently explored\cite{wu:00,chen:07,leptos:09,mino:11,valeriani:11,maggi:17,saintillan:18}. Previous theoretical studies have mainly focused on passive microrheology\cite{gregoire:01,morozov:14,burkholder:17,granek:21} and in particular probing the fluctuation-dissipation relations in active media\cite{maes:14,dalcengio:19,burkholder:19,chaki:18,chaki:19}, or on a constant force active microrheology, either by relying on the low-density description based on the Smoluchowski equation\cite{burkholder:20}, or on numerical simulations for a wide range of suspension densities\cite{reichhardt:15}. Here we use Brownian dynamics simulations to study the active microrheological response of active suspensions. Firstly, we extend the results of reference\cite{reichhardt:15} by subjecting the tracer to a broad range of external forces, and report new results on the resulting nonlinear dependence of the tracer mobility on the driving force. Secondly, we perform an oscillatory active microrheological study of active suspensions. We reveal that the frequency-dependence of tracer mobility can be described by a Lorentzian form in the low frequency domain.  

Our setup is depicted in Fig. \ref{fig:1}: We study a tracer particle of radius $R$, immersed in a two dimensional suspension of active disks, and driven by an external force $\mathbf{F}_\mathrm{e}(t)$. The active disks have a fixed self-propulsion speed $v_\mathrm{A}$, diameter $\sigma$, mobility $\mu_\mathrm{A}$ and they perform persistent motion within a characteristic time $\tau_\mathrm{R} = D^{-1}_\mathrm{R}$, where $D_\mathrm{R}$ is their rotational diffusion constant. They interact among themselves and with the tracer via purely repulsive steric forces. The "bare" mobility of the tracer in a fluid free from active disks is then $\mu_\mathrm{T} = \mu_\mathrm{A}\sigma/2R$. Neglecting hydrodynamic interactions, the motion of the tracer and active disks is described with a set of overdamped stochastic equations, which we solve numerically. The equations of motion and details of the numerical integration scheme are presented in the Methods section.
The suspension of active disks is described by two dimensionless parameters: the P\'eclet number $\text{Pe} = \sigma v_\mathrm{A}/D$ and the total area fraction occupied by disks $\phi = N\sigma^2\pi/(4A)$; here $N$ is the number of disks, $D$ is their translational diffusion constant, and $A$ is the total area of the simulation box.
The P\'eclet number compares the time $t_\mathrm{D} = \sigma^2/D$ it takes an active disk to diffuse its own length with the time $t_\mathrm{S} = \sigma/v_\mathrm{A}$ it takes the disk to swim the same distance. For $\text{Pe} \gg 1$ the active motion dominates over the diffusive transport, while for $\text{Pe} \ll 1$ the disk behaves as a passive particle exhibiting diffusive motion. Throughout this study we set $\phi = 0.12$, so that motiltiy-induced phase separation\cite{filly:12,redner:13,cates:15} does not occur, and select several values of $\text{Pe}$ ranging from the limit of suspensions of passive disks, $\text{Pe} = 0$, to highly active disks, $\text{Pe} = 240$.

The tracer is driven through the active medium with the external force $F_\mathrm{e}(t)$ acting along the $x$-direction with the aim to probe the microrheological properties of the suspension. The collisions with active disks influence the mobility of the tracer in the direction of applied force: instead of having the bare mobility $\mu_\mathrm{T}$ the tracer now acquires an effective mobility $\mu$.

Firstly, we examine the case in which a constant external force, $F_\mathrm{e}(t) = F = \mathrm{const}$ is applied on the tracer. We define the static mobility 
\begin{equation}
\mu = \langle v \rangle/F,
\label{eq:smob}
\end{equation}
where $\langle v \rangle$ is the average speed of the tracer in the $x$-direction in the steady state. We introduce a dimensionless parameter $f = \sigma F/(2Rf_\mathrm{A})$, where $f_\mathrm{A}=v_\mathrm{A}/\mu_\mathrm{A}$ can be interpreted as the force to stop an active disk from moving.
It is easy to see that the tracer driven by a force $f=1$, in a fluid free from disks, moves with a speed $v=v_\mathrm{A}$. It is expected therefore that the speed of the tracer moving in an active suspension should be much smaller than $v_\mathrm{A}$ for $f \ll 1$, and conversely much larger than $v_\mathrm{A}$ for $f \gg 1$. According to our results these two regions are separated by a relatively narrow crossover region located in the vicinity of $f \approx 1$. Indeed, we find that the mobility $\mu$ of the tracer in the active suspension displays a nonlinear dependence on $f$ below this crossover value ($f \lesssim 1$), and it is amplified with respect to the mobility of a tracer driven through the bath of passive disks. On the other hand, for larger forces ($f \gtrsim 1$) the tracer mobilities in passive and active baths match each other, and we find that they are independent of $f$ in this region.

Secondly, we study a tracer driven by a harmonic external force $F_\mathrm{e}(t) = F\sin(\omega t)$, where $\omega$ is the frequency of oscillations, and $F$ now denotes the force amplitude. We find that the velocity $v = \tilde{v} \sin(\omega t + \phi)$ of the tracer, averaged over many independent realizations, oscillates with the same frequency $\omega$ as the driving force, with the phase angle $\phi \approx 0$; on the other hand, the velocity amplitude depends on both the amplitude and the frequency of the driving force, $\tilde{v} = \tilde{v}(F,\omega)$. As a consequence of this the dynamic mobility,
\begin{equation}
\mu(F,\omega) = \tilde{v}(F,\omega)/F,
\label{eq:dmob}
\end{equation}
depends on $\omega$ and $F$ as well. For small values of driving force amplitude ($f \lesssim 1$) we find that the dynamic mobility decreases rather quickly with $\omega$ up to some limiting value of $\omega$ and that our results fit well to a Lorentzian. The Lorentzian spreads out for larger values of $f$. It turns out, however, that for sufficiently large values of $f$ (depending on setup conditions) the mobility $\mu$ changes its behavior -- it grows up rather than falling down with $\omega$. Demonstrating that the dynamic mobility significantly depends on frequency for small driving forces but also on the driving forces itself is the main result of our study.

Recently, it has become common in literature to describe the motion of a tracer in an active bath via a set of effective coarse-grained stochastic equations with a time-independent\cite{angelani:10,maggi:17,knezevic:20} effective friction coefficient. As our results suggest, it is hardly possible to get a satisfactory description in terms of linear stochastic equations in general. They also suggest that in the limit where a linear response is applicable one has to include a memory-mobility kernel\cite{balakrishnan:21,maggi:17,granek:21} instead of simply a constant mobility.

The rest of the article is organized as follows. The Results section is partitioned into two segments presenting constant force and oscillatory force microrheology of suspensions of active disks, respectively. We elaborate the implications of our oscillatory force results and offer our conclusions in the Discussion section. Finally, the Methods section presents equations of motion of the tracer and active disks and provides details of their numerical integration.  

\section*{Results}

In the first part of this section we present our results obtained for the tracer driven through a suspension of active disks by a constant external force, while the second part shows our findings for the tracer guided through the suspension by a harmonic force.

\subsection*{Constant force microrheology}
\begin{figure}[ht]
	\centering
	\includegraphics[width=18cm]{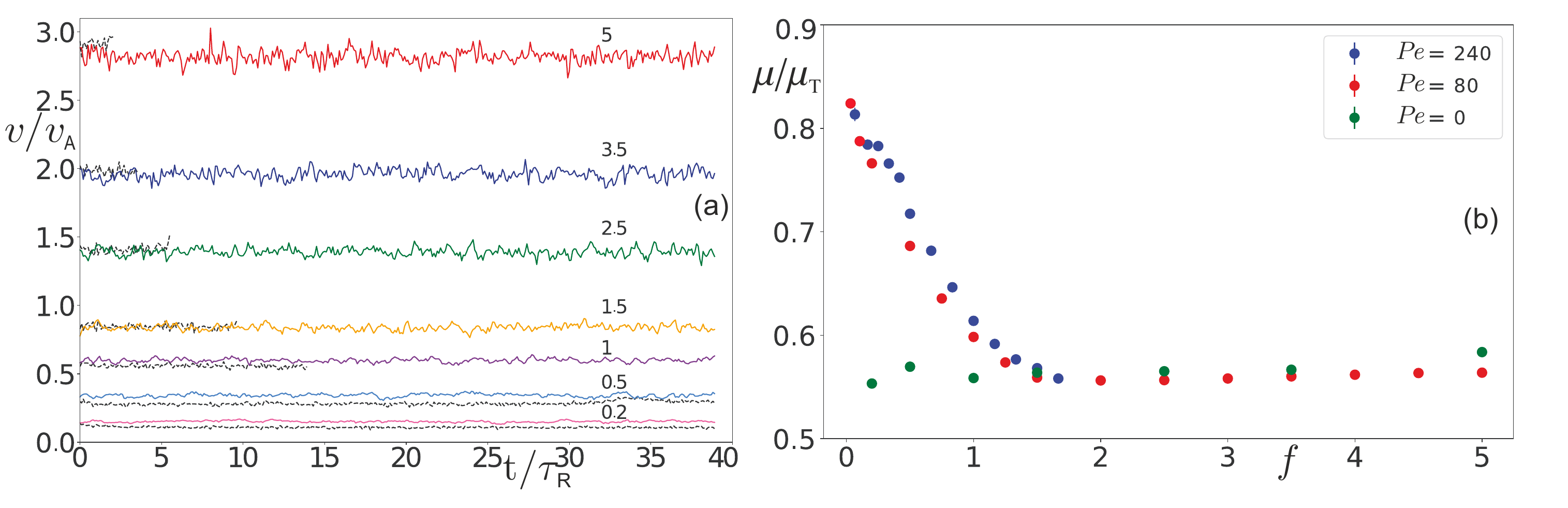}
	\caption{(a) The time fluctuations of the tracer speed $v$ obtained by averaging over 100 independent simulation runs; $v$ is measured in units of active disk speed $v_\mathrm{A}$ and time $t$ in units of active disk reorientation time $\tau_\mathrm{R}$.
	For each value of the driving force $f = 0.2,\, 0.5,\, 1,\, 1.5,\, 2.5,\, 3.5,\, 5$ two sets of lines are displayed: the colored solid lines correspond to the tracer in an active bath with $\text{Pe} = 80$, while the neighboring dashed black lines represent the case of a passive bath, $\text{Pe}=0$. (b) Effective mobility $\mu$ of the tracer measured in units of bare mobility $\mu_\mathrm{T}$ as a function of force $f$ for three different baths described by $\text{Pe} = 0, \, 80, \, 240$. Note, to rescale the driving forces for the passive bath, we use the force scale of the active bath with $\text{Pe} = 80$.}
	\label{fig:2}
\end{figure}

\begin{figure}[h!]
	\centering
	\includegraphics[width=17.5cm]{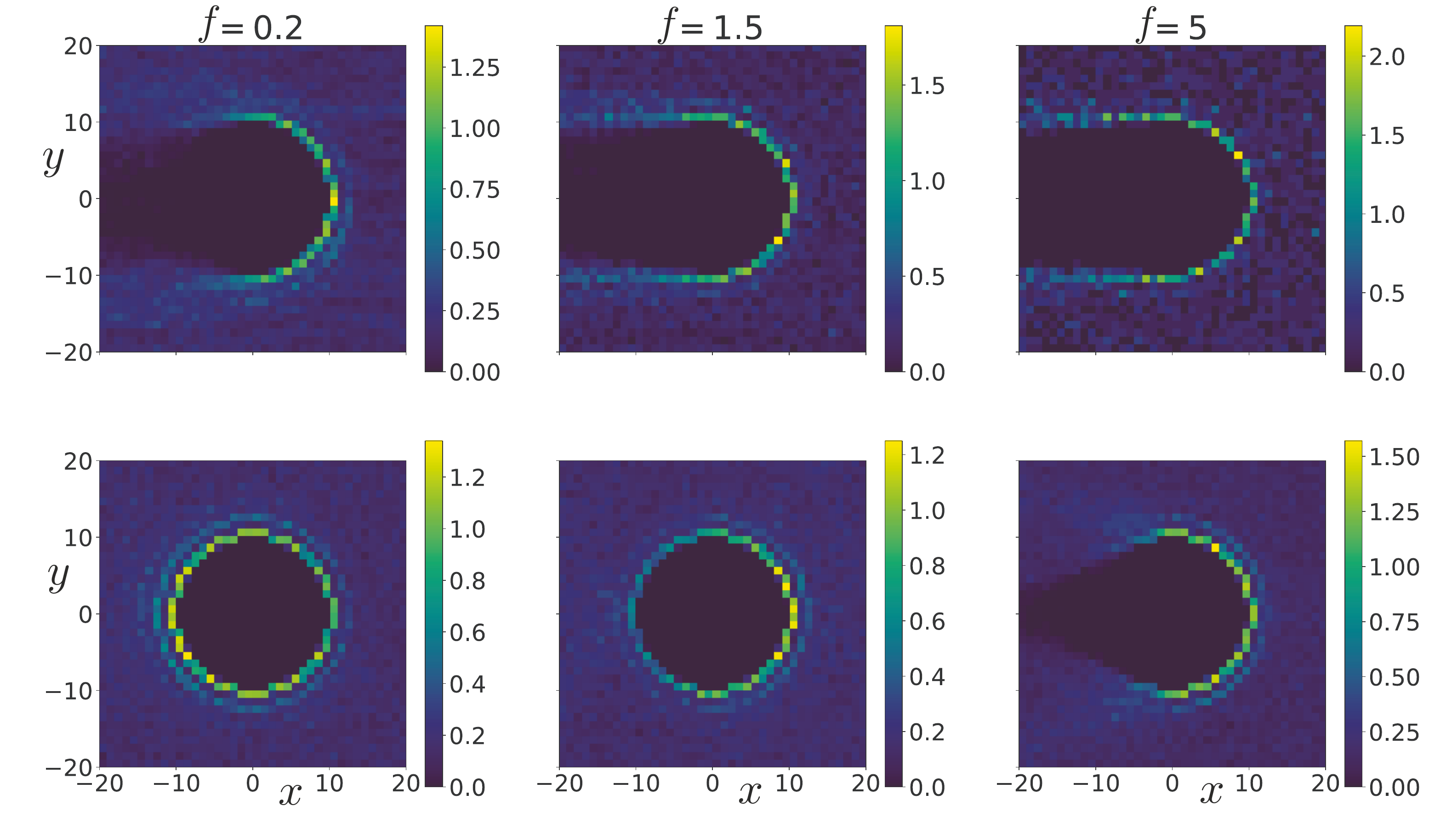}
	\caption{Average density (in units of $\sigma^{-2}$) of disks around the tracer in a passive bath (top row), and in an active bath with $\text{Pe}=80$ (bottom row) for three different external force values $f= 0.2, \, 1.5, \, 5$. In both passive and active bath settings, the force scale of the active bath is used to define the rescaled force $f$.}
	\label{fig:3}
\end{figure}

Taking the average of Eq. (\ref{eq:tracer}) from the Methods section, which describes the tracer dynamics, and using Eq. (\ref{eq:smob}), we find that the effective mobility of the tracer driven by a constant force is
\begin{equation}
\frac{\mu}{\mu_\mathrm{T}} = 1 + \frac{1}{F}\left \langle \sum_{i} F^\mathrm{T}_{ix} \right \rangle,
\label{eq:smobratio}
\end{equation}
where the sum goes over all active disks in the bath and $F^\mathrm{T}_{ix}$ is the $x$-component of the force exerted on the tracer by the $i$-th disk.
We first perform an average over 100 independent simulation runs and then a time average over the steady-state motion of the tracer (see Methods section for details). This double averaging is denoted by $\langle \dots \rangle$. In Fig. \ref{fig:2}(a) (solid lines) we show the temporal fluctuations of the tracer speed $v$, obtained by averaging over independent simulation runs only, for several external force amplitudes $f$. 
One observes a clear increase in tracer speed $v$ with increasing driving force $f$. For the same set of external forces $f$ we also show (black dashed lines) the tracer speed $v$ in a bath of passive disks ($\text{Pe}=0$). For values of driving force $f \lesssim 1$, including the narrow crossover region located around $f \approx 1$, we find that the tracer moves with a larger speed in an active than in a passive bath, and thus has a higher effective mobility, as clearly illustrated in Fig. \ref{fig:2}(b). On the other hand, for larger external forces, $f \gtrsim 1$, the tracer moves with approximately equal speed in the passive and active baths. Thus, in this regime the tracer speed increases linearly with $f$ and the mobility $\mu$ does not depend on $f$. In addition, we find that the mobility $\mu$ versus $f$ data fall roughly on top of each other for baths with different but large activities $\text{Pe}=80$ and $\text{Pe}=240$. This is due to using the rescaled external force $f = \sigma F/(2Rf_\mathrm{A})$, where $f_\mathrm{A}=v_\mathrm{A}/\mu_\mathrm{A}$ is the characteristic force scale of the active bath. For the range of external forces considered, the tracer mobility in the passive bath does not change with $f$, as shown in Fig.\ \ref{fig:2}(b).

To develop some understanding for the observed behavior, in Fig. \ref{fig:3} we plot the average density of disks in the vicinity of the tracer driven through passive and active baths. In passive baths the tracer leaves behind a channel free from disks, which is qualitatively similar for all external forces $f$ that we have considered. For $f \geq 0.2$ the tracer moves so fast that the diffusing passive disks cannot immediately fill the space behind the tracer (see supplementary Movie 1, displaying a tracer driven through a passive bath with $f=0.2$). Thus, the tracer pushes the passive disks in front of it forward, which in turn reduces its mobility with respect to $\mu_\mathrm{T}$, while leaving behind it a trace without disks (cf. Fig. \ref{fig:3}, top row). We find the tracer speed to increase linearly with external driving $f$, leading to a constant mobility $\mu$ for the range of forces considered, Fig. \ref{fig:2}(b). 

In contrast to the passive bath, the tracer moving through the active bath with $\text{Pe}=80$ displays a much richer behavior with changing $f$ in the same range, as the bottom panels in Fig. \ref{fig:3} demonstrate. For large force $f = 5$, the tracer moves faster than the active disks, $v>v_\mathrm{A}$, and consequently leaves a wake behind it; this is illustrated in Movie 2. The shape of the wake is somewhat different compared to the passive bath since the active disks move ballistically into the wake, which hence assumes the shape of a cone (compare top and bottom right panels of Fig. \ref{fig:3}). Nevertheless, the motion of the tracer is qualitatively similar in these two cases. Indeed, the tracer moving with a large speed does not distinguish whether the disks it encounters on its front are passive or active, meaning it has the same effective mobility $\mu$ in both types of baths as demonstrated in Fig.\ \ref{fig:2}(b). As the driving force is gradually reduced towards $f=1$,
by entering the crossover region the speed of the tracer becomes roughly equal to the active disk speed, $v \approx v_\mathrm{A}$, and the active disks start to catch up with the tracer, allowing them to push it from behind. In the density profile of Fig. \ref{fig:3}, middle bottom panel, this manifests itself by the disappearance of the void at the rear side of the tracer. For even lower external force, $f = 0.2$, the active disks move significantly faster than the tracer, and accumulate more at the rear than at the front side of the tracer, left bottom panel in Fig. \ref{fig:3} and Movie 3. Thus, they also push the tracer forward. This explains the increase of the mobility with decreasing $f$ and is investigated in more detail in the following.

\begin{figure}[h!]
	\centering
	\includegraphics[width=17.5cm]{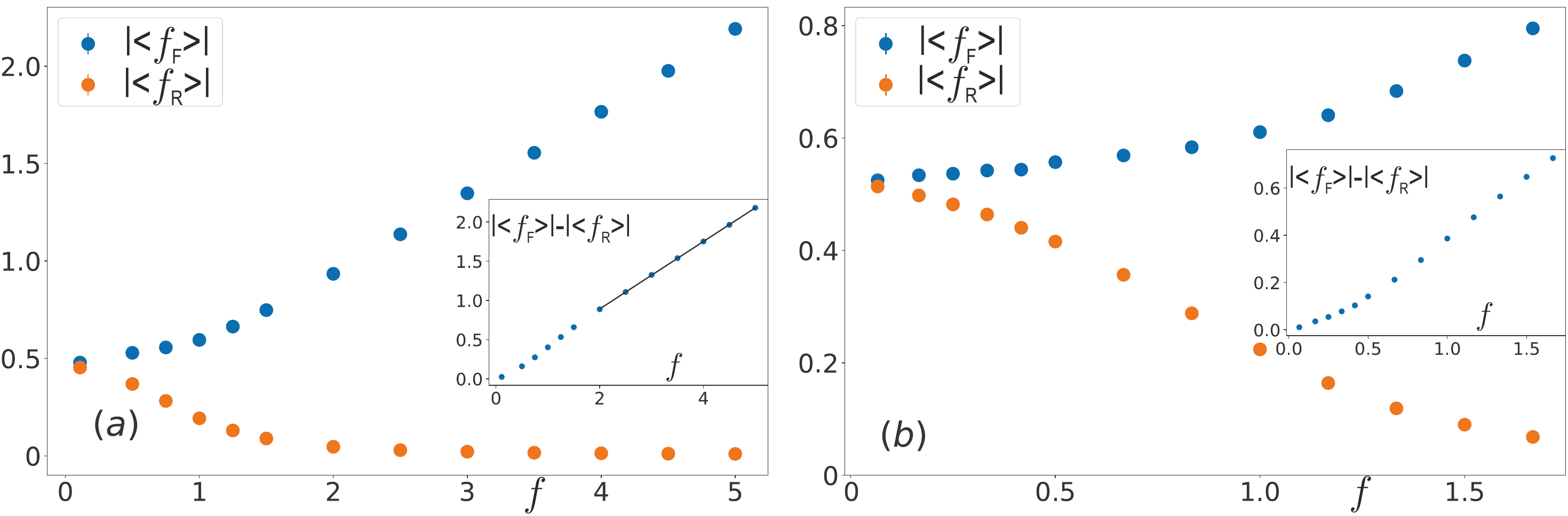}
	\caption{The absolute values of average forces $|\langle f_\mathrm{F} \rangle|$ and $|\langle f_\mathrm{R} \rangle|$ exerted 
	by active disks on the front and rear side of the tra\-cer, respectively, as a function of the external force $f$ for an active bath 
	characterized by (a) $\text{Pe} = 80$ and (b) $\text{Pe}=240$. The insets show the difference $|\langle f_\mathrm{F} \rangle| - |\langle f_\mathrm{R} \rangle|$ versus $f$.}
	\label{fig:4}
\end{figure}

The force $\sum_i F^\mathrm{T}_{ix} = F_\mathrm{F} + F_\mathrm{R}$, with which active disks push against the tracer, can be split into two components $F_\mathrm{F}$ and $F_\mathrm{R}$. They are exerted by the active disks along the $x$-direction on the front half and rear half of the tracer, respectively. We can rewrite Eq. (\ref{eq:smobratio}) to get
\begin{equation}
\frac{\mu}{\mu_\mathrm{T}} = 1 + \frac{\langle f_\mathrm{F} \rangle + \langle f_\mathrm{R} \rangle}{f} 
= 1 - \frac{|\langle f_\mathrm{F} \rangle| - |\langle f_\mathrm{R} \rangle|}{f},
\end{equation}
where we have introduced $f_\mathrm{F} = \sigma F_\mathrm{F} \mu_\mathrm{A}/(2Rv_\mathrm{A})$ and $f_\mathrm{R} = \sigma F_\mathrm{R} \mu_\mathrm{A}/(2Rv_\mathrm{A})$. 
Note that $f_\mathrm{R}$ has the sign of the $x$-component of tracer's velocity, while $f_\mathrm{F}$ has the opposite sign ($\langle f_\mathrm{R} \rangle > 0$, $\langle f_\mathrm{F} \rangle < 0$). In Fig. \ref{fig:4}(a) we plot the absolute values $|\langle f_\mathrm{F} \rangle|$ and $|\langle f_\mathrm{R} \rangle|$ as functions of the external force $f$ for an active bath with $\text{Pe} = 80$. Obviously, the front force has a higher magnitude than the rear force, $|\langle f_\mathrm{F} \rangle| > |\langle f_\mathrm{R} \rangle|$. As can be inferred from Fig. \ref{fig:4}(a), for $f \gtrsim 1$ the force exerted on the tracer front scales linearly with $f$, while the force acting on the rear side of the tracer is vanishingly small, $|\langle f_\mathrm{R} \rangle| \approx 0$. This behavior agrees well with our previous observation of constant tracer mobility $\mu$ for large external force (cf. Fig. \ref{fig:2}(b)). One can also notice that $|\langle f_\mathrm{F} \rangle|$ further decreases with $f$ but in a nonlinear fashion in the region $f \lesssim 1$. On the other hand, $|\langle f_\mathrm{R} \rangle|$ gradually increases with decreasing $f$ in the same region, which corresponds to the onset of active disk accumulation at the rear-side of the tracer as observed in Fig. \ref{fig:3}. Taken together these effects lead to an overall increase of mobility observed for $f \lesssim 1$. It turns out that in the region $f \lesssim 1$ we obtained qualitatively similar behavior for larger values of $\text{Pe}$ number (see Fig. \ref{fig:4}(b) for the case $\text{Pe} = 240$).  

\subsection*{Oscillatory force microrheology}

\begin{figure}[h!]
	\centering
	\includegraphics[width=17.5cm]{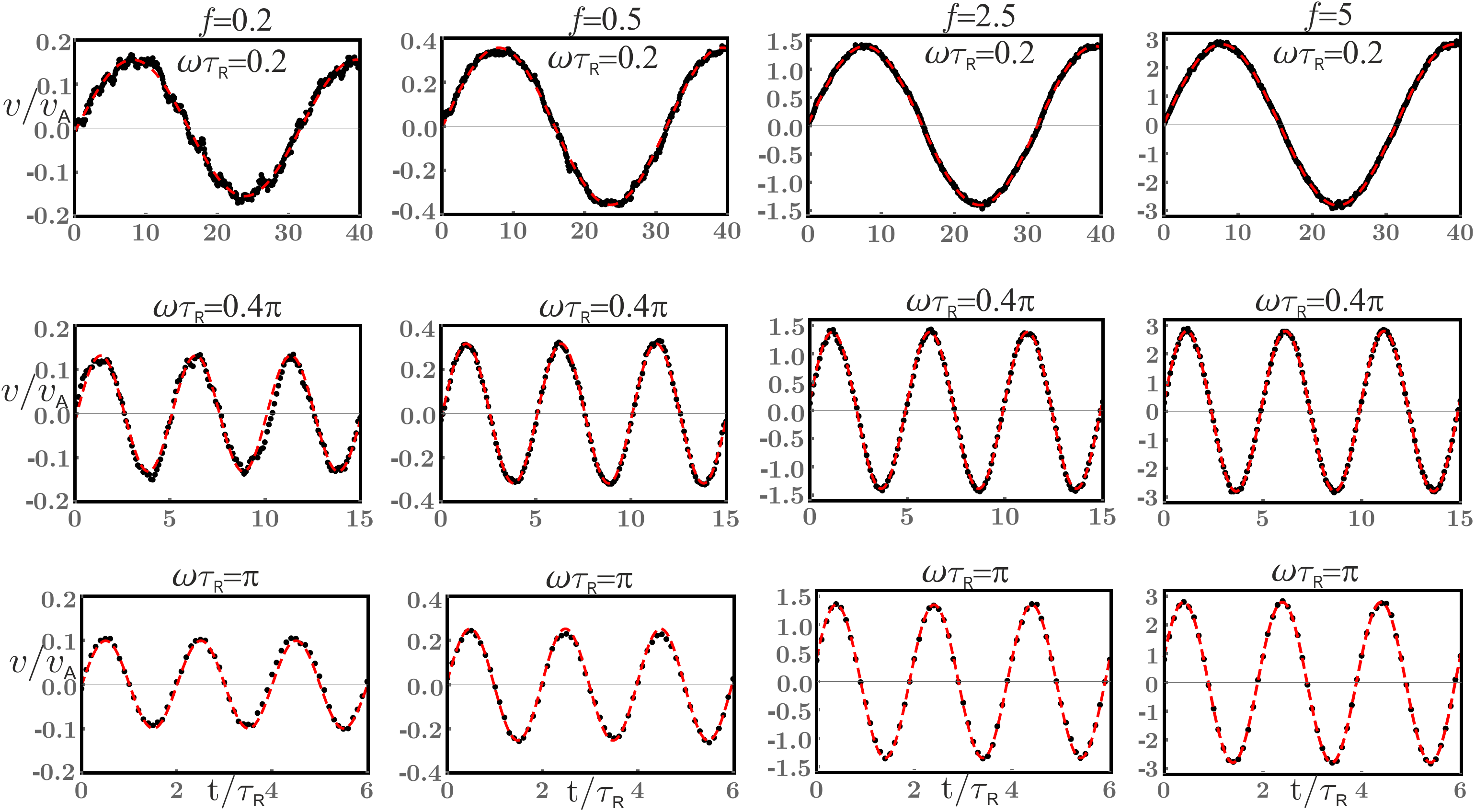}
	\caption{The $x$-component of tracer velocity $v$ measured in units of active disk speed $v_\mathrm{A}$ as a function of time $t$ expressed in units of active disk reorientation time $\tau_\mathrm{R}$. The tracer is driven with an external force $f_\mathrm{e}(t)=f\sin(\omega t)$ through an active bath characterized by $\text{Pe} = 80$. The columns correspond to force amplitudes $f = 0.2, \, 0.5, \, 2.5, \, 5$ while the rows correspond to driving frequencies $\omega\tau_\mathrm{R} = 0.2, \, 0.4\pi, \, \pi$, respectively. Here, black points present simulation data averaged over 100 independent simulation runs for each $t$, while red dashed lines provide the best fits of the data to the form 
	$\tilde{v} \sin(\omega t + \phi)$.}
	\label{fig:5}
\end{figure}

Having established how the tracer mobility depends on a static external force $f$ in an active suspension, we turn to the case of an oscillatory external force. We subject the tracer to a sinusoidal force $f_\mathrm{e}(t) = f \sin(\omega t)$ acting along the $x$-axis where $f$ denotes the force amplitude and $\omega$ the frequency. For a bath with $\text{Pe} = 80$ the time profiles of the tracer velocity $v$ along the $x$-axis for some representative values of $f$ and $\omega$ are shown in Fig. \ref{fig:5}. The time evolution of the velocity $v(t)$ has been calculated by averaging over 100 independent simulation runs. The velocity profiles were then fitted to a simple form $v(t) = \tilde{v} \sin(\omega t + \phi)$, with the velocity amplitude $\tilde{v}$ and phase shift $\phi$ being the fit parameters. We have been able to get very good fits with $\phi \approx 0$ for all amplitudes $f$ and frequencies $\omega$ examined in our study, which clearly has to be assigned to the active or nonequilibrium nature of the bath particles performing an overdamped motion.
On the other hand, we find that the velocity amplitude depends both on the amplitude and frequency of the driving force, $\tilde{v}=\tilde{v}(f, \omega)$. As one can infer from Fig. \ref{fig:5}, for a fixed $\omega$ the amplitude $\tilde{v}$ grows quickly with $f$. For a fixed $f$, $\tilde{v}$ decreases as a function of $\omega$ for relatively small values of $f$ (e.g. $f=0.2, \, 0.5$) while it depends only weakly on $\omega$ for larger values of $f$ (e.g. $f=2.5, \, 5$). Some qualitative features of tracer motion can be seen in Movie 4 (for $f=0.5$ and $\omega\tau_{R}=0.6\pi$) and in Movie 5 (for $f=2.5$ and $\omega\tau_\mathrm{R}=\pi$).

\begin{figure}[h!]
	\centering
	\includegraphics[width=15cm]{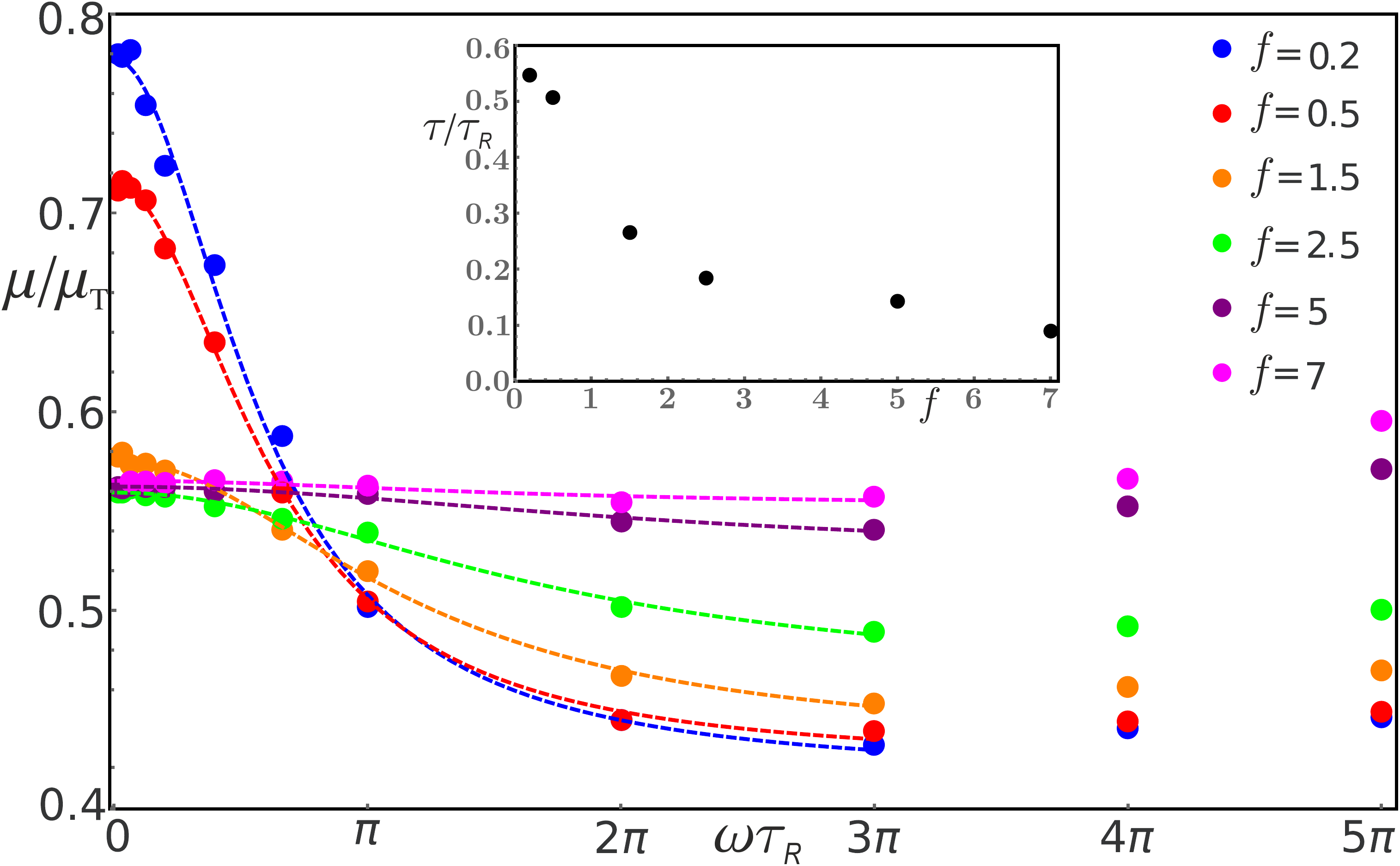}
	\caption{The dynamic mobility of the tracer $\mu(\omega)$ measured in units of bare mobility $\mu_\mathrm{T}$ as a function of the frequency $\omega$ expressed in units of $\tau_\mathrm{R}^{-1}$. The mobility $\mu(\omega)$ is measured for several driving force amplitudes $f=0.2, \, 0.5, \, 1.5, \, 2.5, \, 5, \, 7$. The dashed lines provide best fits of the mobility data to the Lorentz form $\mu_{\infty} + \mu_0/[1 + (\omega \tau)^2]$ for $\omega\tau_{R}$ up to $3\pi$, while in the inset we show values of the fit parameter $\tau$ as a function of $f$.}
	\label{fig:6}
\end{figure}

The variation of the dynamic mobility $\mu(\omega)$ with frequency $\omega$ in a suspension having $\text{Pe} = 80$ is calculated using Eq. (\ref{eq:dmob}) and presented in Fig. \ref{fig:6}. Depending on the magnitude of the amplitude $f$, we distinguish two qualitatively different regimes.  
In the case of small and moderate amplitudes ($f \lesssim 7$ for the case $\text{Pe}=80$) we find that $\mu(\omega)$ decreases with $\omega$ up to approximately $\omega\tau_\mathrm{R} = 3\pi$. In particular, in this region $\mu(\omega)$ changes quite quickly for the case of small amplitudes ($f = 0.2, \, 0.5$), but the change slows down as $f$ increases gradually ($f = 1.5, \, 2.5, \, 5, \, 7$). Nevertheless, in this region all of our mobility data sets fit quite well to the shifted Lorentz form: $\mu_{\infty} + \mu_0/[1 + (\omega \tau)^2]$, where the fit parameters $\mu_{\infty}$, $\mu_0$ and $\tau$ depend on the amplitude $f$. We find that $\tau^{-1}$, which provides a measure of the characteristic width of the Lorentz form, increases with $f$ (see Fig. \ref{fig:6}). For larger frequencies ($\omega \tau_\mathrm{R} > 3\pi$) it seems that $\mu$ slowly increases with $\omega$, somewhat faster for larger values of $f$.

\begin{figure}[h!]
	\centering
	\includegraphics[width=17.5cm]{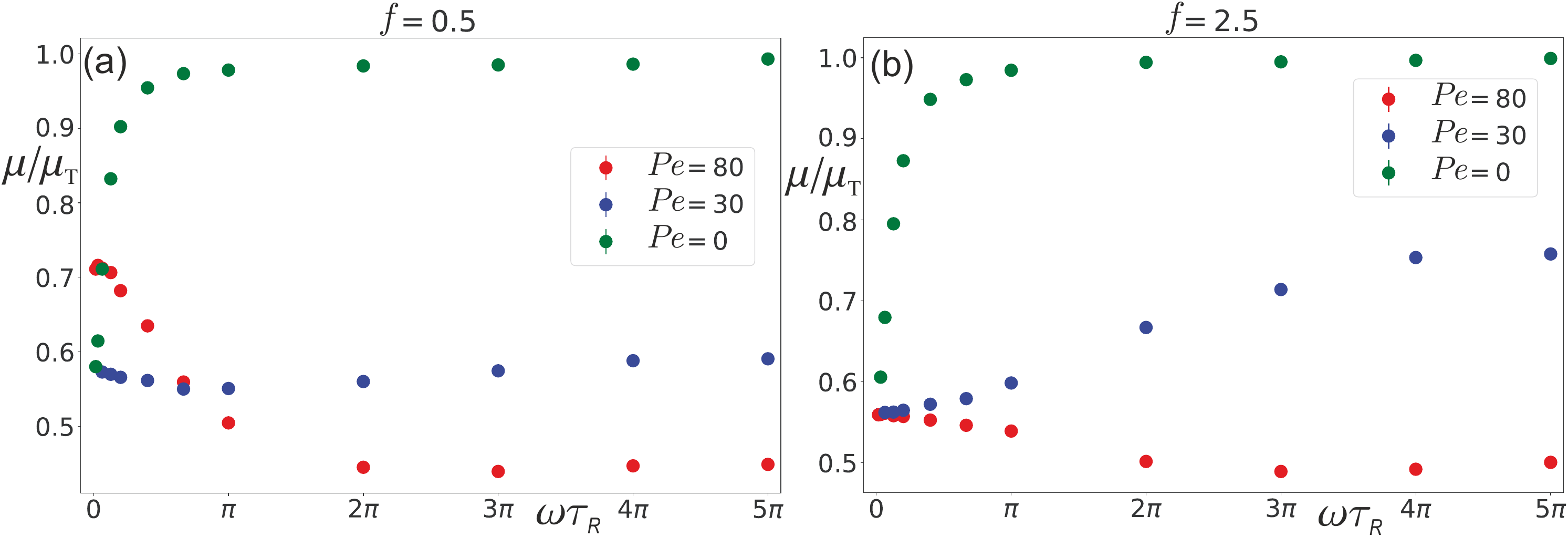}
	\caption{The tracer mobility $\mu(\omega)$ as a function of the frequency $\omega$ for the case of two active baths $\text{Pe} =  30, \, 80$, and a passive bath $\text{Pe}=0$. The tracer is driven by a harmonic force of amplitude (a) $f=0.5$, (b) $f=2.5$.}
	\label{fig:7}
\end{figure}

As one can also notice from Fig. \ref{fig:6}, the overall shape of the curve $\mu(\omega)$ changes considerably by increasing $f$ (compare e.g. the cases $f=0.2$ and $f=7$). It would be interesting therefore to explore the regime in which the magnitude of the driving force significantly exceeds the self-propulsion force of active disks. It is expected, namely, that in the limit of large force ($f \gg 7$ for $\text{Pe}=80$), the dynamic mobility $\mu(\omega)$ should be alike to that one finds in the passive bath (note that in the passive case $\mu(\omega)$ is an increasing rather than decreasing function of $\omega$, as illustrated in Fig. \ref{fig:7}). It is however rather difficult to perform numerical simulations in this limit because in that case one has to deal with a very small simulation time step $\Delta t$. Instead, one can consider a tracer driven by a force of smaller amplitude $f$, but immersed in a bath having a lower $\text{Pe}$ number. For $\text{Pe}=30$, two examples are given in Fig. \ref{fig:7} for amplitudes $f=0.5$ and $f=2.5$. One can see that in the case $f=2.5$, the mobility $\mu(\omega)$ is a nondecreasing function of $\omega$. For $f=0.5$ the mobility decreases in the low frequency domain, and after that, in the region of larger frequencies $\mu(\omega)$ increases.
However, interestingly the variation in the probed frequency range is rather small. For the sake of comparison, in Fig. \ref{fig:7} we also 
present the mobility for the bath with $\text{Pe}=80$, as well as for the case of a passive bath ($\text{Pe}=0$). In contrast to the case of active bath,the mobility of the tracer in a passive bath jumps up quickly to a plateau value $\mu = \mu_\mathrm{T}$ in the low frequency region. This can be explained by the fact that a tracer oscillating with a higher frequency moves practically in a region without obstacles, since the time $t_\mathrm{R} = R^2/4D = 12 \tau_\mathrm{R}$ it takes a passive disk to diffuse the tracer radius $R$ is larger than the period $T = 2\pi/\omega$ of tracer oscillations. Conversely, when $t_\mathrm{R}<T$ the disks have enough time to diffuse into the wake created by the moving tracer, and thus they act as obstacles, which leads to a decrease of the mobility. Then the crossover region between these two regimes can be roughly estimated by using the condition $t_\mathrm{R} \approx T$, which gives $\omega\tau_\mathrm{R} \approx \pi/6$. As one can verify, this estimate indeed falls into the crossover region, see Fig. \ref{fig:7}.   

\section*{Discussion}

In this article we studied the microrheological properties of 2d suspensions of active disks. For this purpose we considered the motion of a tracer particle immersed in the active suspension. The tracer was driven either by a constant or an oscillatory force. In the case of a constant driving force, we found that the mobility of the tracer significantly decreases with growing magnitude of the external force $f$ for weaker strengths
($f \lesssim 1$), while it approaches an approximately constant value for stronger forces ($f \gtrsim 1$) [see Fig. \ref{fig:2}(b)], which agrees with the tracer mobility in a suspension of passive particles. We stress that we did not observe a plateau in the mobility in the region of small $f$ values, meaning that a linear response approach is not applicable in our system. Thus, the active suspension exhibits a highly nonlinear behavior for weak forces, $f \lesssim 1$. 
As a consequence of this, a simple coarse-grained description\cite{wu:00,angelani:10,maggi:17,knezevic:20} of tracer dynamics in active suspensions in terms of an effective linear stochastic equation is not applicable in our case. It is interesting to note that in the region of very small forces ($f \lesssim 0.01$) a mobility plateau has been observed in active suspensions in the low density limit\cite{burkholder:20}. This is in contrast with our observations for active suspensions, at least for $\text{Pe}\ge 80$ and the range of $f$ values we explored so far. We note
that much more numerical efforts are needed to get a reliable estimate of the mobility for very small values of $f$, due to very large fluctuations of the tracer velocity. For this reason the existence of a region in which the system displays a linear response remains open for now. 

Importantly, we find that the dynamic mobility of a tracer driven by an oscillatory force is a quite complex quantity due to the ability of the active disks to push against the tracer. The mobility shows a strong frequency dependence up to moderate force amplitudes ($f \lesssim 2.5$ in the case of $\text{Pe}=80$). In the range of frequencies $\omega\tau_{R}\le 3\pi$ the mobility decreases with $\omega$ and its frequency dependence can be well described by a Lorentzian that spreads out with $f$, as presented in Fig. \ref{fig:6}. Neglecting the weak increase of the mobility observed at high frequencies, this means that the tracer motion can be described by an effective Langevin equation with an exponential memory-mobility kernel but only provided that the system's response is linear for very small $f$. Our aim in the future is to provide a more detailed numerical study of the system in this region.

In Ref.\ \cite{chen:07} the active microrheology experiments performed in a bacterial bath do not show any frequency dependence in the dynamic mobility. We attribute this to a bacterial density of $\phi = 3 \cdot 10^{-3}$, which is much smaller than $\phi = 0.12$ used in our work.
Thus, in this article we demonstrate that at moderate densities the dynamic mobility depends on frequency and shows a highly nonlinear 
behavior in the driving force. With our simulation work we provide a clear orientation where to search for this behavior in the available 
parameter space when performing active microrheology experiments of active suspensions of artifical or biological microswimmers.

\section*{Methods}

We study a system of $N$ interacting active disks in 2 dimensions which move with a constant speed $v_\mathrm{A}$ and have mobility $\mu_\mathrm{A}$. Their translational and rotational dynamics are described by a set of coupled overdamped Langevin equations
\begin{eqnarray}
\dot{\mathbf{r}}_i &=& v_\mathrm{A} \mathbf{n}_i + \mu_\mathrm{A} \left ( \sum_{j \neq i} \mathbf{F}^{\mathrm{A}}_{ij} - \mathbf{F}^\mathrm{T}_i \right ) + \sqrt{2D}\bm{\xi}_i, \label{eq:active1} \\
\dot{\theta}_i &=& \sqrt{2D_{\mathrm{R}}} \eta_i. \label{eq:active2}
\end{eqnarray}
Here $\mathbf{r}_i$ denotes the position vector and $\mathbf{n}_i = (\cos\theta_i, \sin\theta_i)$ the unit orientation vector of the $i$-th disk. The disks are subjected to Gaussian noises of zero mean, $\langle \bm{\xi}_i(t) \rangle = 0$ and $\langle \eta_i(t) \rangle = 0$, and unit variance, $\langle \xi^\alpha_i(t) \xi^\beta_j(t) \rangle = \delta_{\alpha \beta}\delta_{ij}\delta(t-t')$ and $\langle \eta_i(t) \eta_j(t') \rangle = \delta_{ij} \delta(t-t')$, and have translational diffusivity $D$ and rotational diffusivity $D_\mathrm{R}$. They interact among themselves via purely repulsive pairwise forces $\mathbf{F}^{\mathrm{A}}_{ij} = - \nabla_{\mathbf{r}_i} V_\sigma(\mathbf{r}_i - \mathbf{r}_j)$ which stem from the Weeks--Chandler--Andersen (WCA) potential
\begin{equation}
V_\sigma(\mathbf{r}) = \left\{
\begin{array}{ll}
4 \varepsilon \left [ \left (\frac{\sigma}{|\mathbf{r}|} \right )^{12} - \left (\frac{\sigma}{|\mathbf{r}|} \right )^{6}\right ] + \varepsilon, & \quad |\mathbf{r}| \leq 2^{1/6} \sigma, \\
0, & \quad |\mathbf{r}| > 2^{1/6} \sigma.
\end{array}
\right.
\label{eq:WCA}
\end{equation}
The potential introduces a characteristic distance $\sigma$, where it assumes the strength $\varepsilon$. One can roughly interpret this distance as the diameter of the active disks. The force $-\mathbf{F}^\mathrm{T}_i$ is connected to the passive tracer, which we introduce now.

A passive tracer disk of radius $R$ and mobility $\mu_\mathrm{T} = \mu_\mathrm{A}\sigma/2R$ is immersed in the active bath and driven by an external force $\mathbf{F}_\mathrm{e}(t) = F_\mathrm{e}(t) \mathbf{e}_x$. We investigate two different settings: a time-independent driving force $F_\mathrm{e}(t)=F$ and a harmonic force $F_\mathrm{e}(t) = F\sin(\omega t)$, where $\omega$ is the oscillation frequency. The position vector $\mathbf{r}_T$ of the tracer changes according to
\begin{equation}
\dot{\mathbf{r}}_{\mathrm{T}} = \mu_\mathrm{T} \left (\sum_{i} \mathbf{F}^{\mathrm{T}}_i + \mathbf{F}_\mathrm{e}(t) \right ). 
\label{eq:tracer}
\end{equation}
The $i$-th active disk pushes the tracer with a force $\mathbf{F}^{\mathrm{T}}_i = - \nabla_{\mathbf{r}_{\mathrm{T}}} V_d(\mathbf{r}_{\mathrm{T}} - \mathbf{r}_i)$, with $d=R+\sigma/2$ the characteristic interaction distance between them, and $V_d$ is also a WCA potential as introduced in Eq. (\ref{eq:WCA}).
Therefore, a force of equal magnitude and opposite sign is included into Eq. (\ref{eq:active1}). We are interested in measuring the effective mobility $\mu$ of the tracer along the $x$-direction, which is primarily influenced by its interactions with the surrounding active disks. This allowed us to neglect thermal motion of the tracer in Eq. (\ref{eq:tracer}).

We use $\sigma$ as the unit of length, the characteristic reorientation time of an active disk $\tau_\mathrm{R} = D^{-1}_\mathrm{R}$ as the unit of time, and $k_\mathrm{B}T$ as the unit of energy. Assuming that $D=D_\mathrm{R}\sigma^2/3$, the mobility of an active disk takes the value
$\mu_\mathrm{A} = 1/3 \,\,\, \sigma^2/(\tau_\mathrm{R}k_\mathrm{B}T)$. Disks are placed in a box of area $A$ and are subjected to periodic boundary conditions. The active bath is described by two dimensionless parameters: the P\'eclet number $\text{Pe} = \sigma v_\mathrm{A}/D$ and the area packing fraction of disks $\phi = N\sigma^2\pi/(4A)$.

We set $N=10000$, $\phi = 0.12$, $\varepsilon/k_\mathrm{B}T=100$, $R/\sigma = 4$ and consider baths characterized by various $\text{Pe}$ numbers, including the limiting case of passive disks, $\text{Pe}=0$. The amplitude $F$ and frequency $\omega$ of the external force are varied according to the protocol explained in the Results section. The Eqs. (\ref{eq:active1}) -- (\ref{eq:tracer}) are numerically integrated using an Euler scheme with a time step of $\Delta t = 10^{-5} \tau_\mathrm{R}$. We first ensure that the tracer in the active bath has reached a steady state in the absence of a driving force. Then the external force is introduced, and data are collected in simulations running over a time period of $40\tau_\mathrm{R}$ and averaged over 100 independent simulation runs, unless stated otherwise.

\section*{Acknowledgements}

MK acknowledges financial support from TU Berlin through a visiting lectureship.

\section*{Author contributions statement}

MK and HS designed research. LEAP and MK performed simulations and analyzed data. MK and HS wrote the article.

\section*{Competing interests}

The authors declare no competing interests.

\section*{Additional information}

\noindent \textbf{Correspondence} and request for materials should be sent to MK.

\end{document}